\documentclass[chicago, iop, apj]{emulateapj}

\usepackage{natbib}
\usepackage{bm}
\usepackage{times}
\usepackage{color}

\def\cii    {[C{\sc\,ii}]}

\newcommand{\pcb}{PIXIE\,$\times$\,BOSS}
\newcommand{\pixiefwhm}{$1.65^\circ$\, FWHM}
\newcommand{\sbgas}{\bar S_{\rm line} b_{\rm line}}
\newcommand{\sbcii}{\bar S_{\rm CII} b_{\rm CII}}
\newcommand{\sbrgas}{\bar S_{\rm line} b_{\rm line} r_\ell}
\newcommand{\mjysr}{\,{\rm MJy}\,{\rm sr}^{-1}}
\newcommand{\kjysr}{\,{\rm kJy}\,{\rm sr}^{-1}}

\begin{document}

\author{
Eric~R.~Switzer\altaffilmark{1}
}
\altaffiltext{1}{NASA Goddard Space Flight Center, Greenbelt, MD, USA}
\email{eric.r.switzer@nasa.gov}
\shortauthors{Switzer}

\accepted{March 8, 2017}

\shorttitle{Intensity mapping with CMB spectral surveys}

\title{Tracing the cosmological evolution of stars and cold gas with CMB spectral surveys}

\begin{abstract}
A full account of galaxy evolution in the context of $\Lambda$CDM cosmology requires measurements of the average star-formation rate (SFR) and cold gas abundance across cosmic time. Emission from the CO ladder traces cold gas, and \cii\ fine structure emission at $158\,\mu{\rm m}$ traces the SFR. Intensity mapping surveys the cumulative surface brightness of emitting lines as a function of redshift, rather than individual galaxies. CMB spectral distortion instruments are sensitive to both the mean and anisotropy of the intensity of redshifted CO and \cii\ emission. Large-scale anisotropy is proportional to the product of the mean surface brightness and the line luminosity-weighted bias. The bias provides a connection between galaxy evolution and its cosmological context, and is a unique asset of intensity mapping. Cross-correlation with galaxy redshift surveys allows unambiguous measurements of redshifted line brightness despite residual continuum contamination and interlopers. Measurement of line brightness through cross-correlation also evades cosmic variance and suggests new observation strategies. Galactic foreground emission is $\approx 10^3$ times larger than the expected signals, and this places stringent requirements on instrument calibration and stability. Under a range of assumptions, a linear combination of bands cleans continuum contamination sufficiently that residuals produce a modest penalty over the instrumental noise. For PIXIE, the $2 \sigma$ sensitivity to CO and \cii\ emission scales from $\approx 5\times 10^{-2}\kjysr$ at low redshift to $\approx 2\kjysr$ by reionization.
\end{abstract}

\section{Introduction}
\label{sec:intro}

Stars form in condensations of cold ${\rm H}_2$ gas \citep{2012ARA&A..50..531K}. The average abundance and cosmological evolution of this gas are poorly constrained \citep{2013ARA&A..51..105C}. Additional measurements will improve our understanding of star-formation efficiency and the divergence of the star-formation rate (SFR) relative to the continued growth of dark matter structure. Cold, molecular gas is traced well by a ladder of CO emission lines at $\approx 115 J_i$\, GHz (for the transition $J_i$ to $J_f=J_i-1$). The $^2P_{3/2} \rightarrow ^2P_{1/2}$ fine structure transition in \cii\ at $158\,\mu{\rm m}$ traces the SFR. CO and \cii\ are excellent diagnostics of galaxy evolution.

Surveys of line emission from individual objects must account for Poisson and cosmic variance, and for any effects due to the selection of the sample. One- and two-point statistics \citep[e.g.][]{2010MNRAS.409..109G, 2013ApJ...772...77V} of continuum emission have the potential to reach to lower flux, but lack precise redshift information. Intensity mapping \citep{1979MNRAS.188..791H} is a hybrid of individual line emission searches and two-point studies of the dust continuum. It surveys the sum of all line radiation as a function of redshift, and requires angular resolution to reach cosmological scales, but not to resolve individual sources. It directly and efficiently measures the first and second moments of the luminosity function from all emitting objects, potentially performing an unbiased census from reionization to the present. Intensity mapping is uniquely sensitive to the line luminosity-weighted bias of emitting gas. Statistical analysis for the power spectrum can average over all modes in the survey, yielding high sensitivities. 

Intensity mapping techniques have provided several informative constraints on galaxy evolution since reionization. The COPSS-II survey \citep{2016ApJ...830...34K} has used SZA to determine the amplitude of the power spectrum of CO brightness fluctuations at $z\approx3$ as $3.0^{+1.3}_{-1.3} \times 10^3\,\mu{\rm K}^2 (h^{-1}\,{\rm Mpc})^3$, which they interpret as $\bar \rho_{H_2}(z=3)\,=\,1.1_{-0.4}^{+0.7} \times 10^{8} M_{\odot}\,{\rm Mpc}^{-3}$. \citet{2016MNRAS.457.3541C} use BOSS to measure the mean surface brightness of redshifted Ly$\alpha$ in cross-correlation between quasars and spectra, finding mean surface brightness $\bar S_\alpha$ multiplied by bias $b_\alpha$ of $\bar S_\alpha (b_\alpha /3) = (3.9 \pm 0.9) \times 10^{-21}\,{\rm erg}\,{\rm s}^{-1}\,{\rm cm}^{-2}\,\AA^{-1}\,{\rm arcsec}^{-2}$ across $z=2-3.5$, a factor of $\sim30$ higher than previously expected. \citet{2013MNRAS.434L..46S} use the GBT to measure $21$\,cm auto- and cross-power with WiggleZ to constrain neutral hydrogen abundance multiplied by bias at $z\sim 0.8$ as $\Omega_{\rm HI} b_{\rm HI} = 0.62_{-0.15}^{+0.23} \times 10^{-3}$.

The PIXIE \citep{2014SPIE.9143E..1EK} and PRISM \citep{2014JCAP...02..006A} missions propose to make deep spectral maps to search for CMB spectral distortions. These data volumes would have a unique sensitivity to CO and \cii\ emission. PIXIE's sensitivity of $\approx 1\kjysr$ per $1^\circ \times 1^\circ \times 15\,{\rm GHz}$ voxel with a \pixiefwhm\ beam would probe CO and \cii\ mean emission and fluctuations in the linear regime of large-scale structure. 

Deep spectral maps contain all sources of radiation in each voxel in addition to the lines of interest: galactic emission, extragalactic thermal emission (cosmic infrared background, CIB), and lines from other redshifts (interlopers). A linear combination of maps can strongly suppress continuum contamination, but the degree of suppression may be limited by instrumental calibration and stability. Residuals after cleaning these sources of contamination additively bias both the auto-power and the spectral monopole from a given emission line. Cross-correlation can unambiguously determine the line surface brightness under contamination from uncorrelated residuals \citep[e.g][]{2013ApJ...763L..20M, 2016MNRAS.457.3541C}.

Cross-correlation finds coherence between galaxy count-density $n$ and line surface brightness $S$ through underlying cosmological overdensity $\delta = (\rho - \bar \rho)/ \bar \rho$ as $S \leftarrow \delta \rightarrow n$. The cross-correlation tracks all emitting gas, not only stacked emission from the galaxies in the spectroscopic survey. Line brightness determined through the cross-power does not have a cosmic variance or require a detailed model of the galaxy power spectrum. 

\citet{2008A&A...489..489R} and \citet{2014MNRAS.443.3506B} have developed 2D anisotropy statistics for CO, and \citet{2013ApJ...768...15P} have considered cross-correlation with broadband CMB surveys. \citet{2016MNRAS.458L..99M} and \citet{2016ApJ...833..153S} further calculate the global signal from CO and \cii\ in the context of PIXIE. \citet{2014ApJ...793..116U} considered high-resolution surveys for \cii\ at intermediate redshift. This paper combines several threads and describes CO and \cii\ anisotropy in cross-correlation at large angular scales probed by CMB spectral surveys such as PIXIE. The multi-tracer approach \citep{2009JCAP...10..007M, 2011MNRAS.416.3009B} for evading cosmic variance impacts intensity mapping survey planning. 

\section{Line emission and observational parameters}
\label{sec:galaxyconnection}

\subsection{The CO ladder and \cii\ emission}

${\rm H}_2$ transitions are poor tracers of star-forming regions. The CO molecule is present in similar environments and has a lowest $J=1-0$ excitation of $h \nu / k_B = 5.5$\,K. CO has a regular ladder at $\approx 115 J_i$\, GHz which is excited in critical densities from $2\times 10^3\,{\rm cm}^{-3}$ to $1\times 10^6\,{\rm cm}^{-3}$, from $J=1-0$ to $J=10-9$, respectively \citep{2013ARA&A..51..105C}. The spectral line energy distribution (SLED) of the relative intensity of the CO ladder depends on $n_{H_2}$ and the kinetic temperature, and thus can be used to trace those quantities. The brightness of the $J=1-0$ transition directly maps to the ${\rm H}_2$ abundance \citep{2013ARA&A..51..207B}. Toward higher redshifts ($z \gtrsim 2$), galaxies may lack metals or dust, leading to a predicted evolution in the CO relation to a greater \citep{1997A&A...328..471I} or lesser \citep{2009ApJ...702.1321O, 2011MNRAS.412..337G} degree.

\begin{figure}
\includegraphics[scale=0.6]{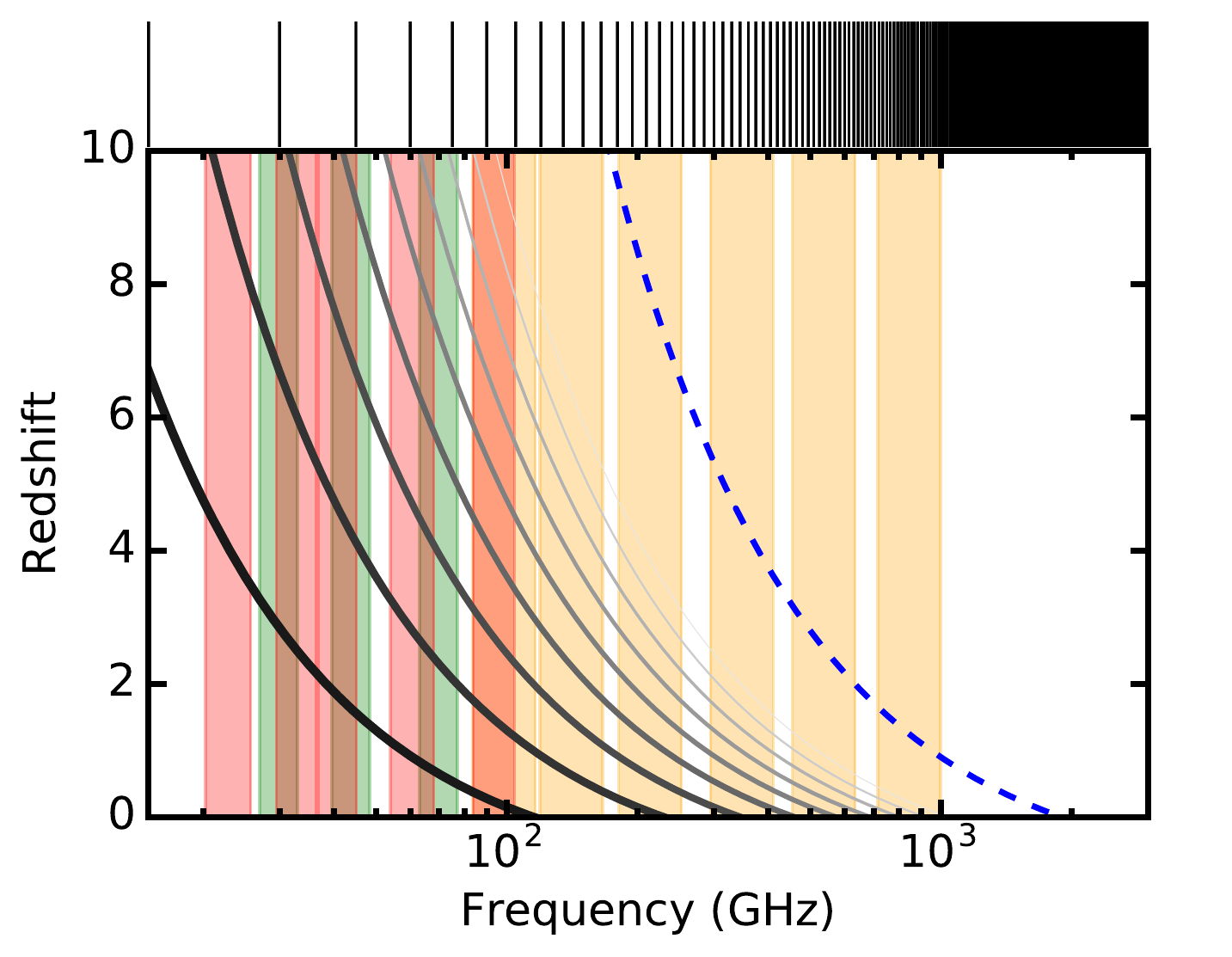}
\caption{Visibility of the CO (black for $J=1-0$ to light gray moving up the ladder) and \cii\ (dashed blue) lines as a function of redshift. WMAP (red) and Planck (LFI in green and HFI in orange) have sensitivity in wide, photometric bands, over which density contrast is washed out. PIXIE's bands (black edges along the top) sample \cii\ and CO from the present to reionization.}
\label{fig:line_visibility}
\end{figure}

The $158\,\mu{\rm m}$ ($1900$~GHz) $^2P_{3/2} \rightarrow\,^2P_{1/2}$ fine structure transition of singly ionized carbon \cii\ is the brightest Far-IR cooling line, emitting $0.5\%-1\%$ of the total Far-IR luminosity \citep{1997ApJ...491L..27M, 1998ApJ...504L..11L, 2010ApJ...724..957S, 2011ApJ...728L...7G}. \cii\ and has shown promise as a tracer of the SFR \citep{2013ARA&A..51..105C}. Given its $11.6$\,eV ionization energy, \cii\ exists in almost all phases of star forming regions \citep{2013A&A...554A.103P, 2014A&A...570A.121P}, but preferentially in warm and dense photo-disassociation regions on the UV-illuminated edges of molecular clouds. \citet{2011MNRAS.416.2712D} report a relation between \cii\ luminosity and the SFR in local, late-type star forming galaxies
\begin{equation}
{\rm SFR}[M_\odot\,{\rm yr}^{-1}] = \frac{(L_{\rm CII} [{\rm erg}\,{\rm s}^{-1}])^{0.983}}{1.028 \times 10^{40}}.
\label{eqn:sfrrelation}
\end{equation}
\cii\ provides an alternative to UV and thermal dust tracers of the SFR: it has different extinction properties than UV, and in contrast to continuum dust tracers, it provides a discrete redshift. Fig.\,\ref{fig:line_visibility} shows PIXIE, Planck, and WMAP bands compared to redshifted CO and \cii\ emission frequencies.

Cosmological predictions \citep[e.g.][]{2016ApJ...817..169L} for the surface brightness depend on a chain from (1) the formation of dark matter halos, (2) the SFR within halos, (3) the implied IR luminosity, (4) a relation between IR luminosity and line luminosity. Each relation exhibits some halo-to-halo scatter. Empirically, the surface brightness of the CO and \cii\ lines is tied \citep{2014ApJ...793..116U} to the IR luminosity function $\Phi$ and the line luminosity as a function of IR luminosity $L_{\rm line}(L_{\rm IR})$ as
\begin{equation}
\label{eqn:lirsb}
\bar S = \int d \log L_{\rm IR} \Phi(L_{\rm IR}) \frac{L_{\rm line}(L_{\rm IR})}{4 \pi D_L^2} y D_A^2.
\end{equation}
where $D_L$ and $D_A$ are the luminosity and angular diameter distance, and $y=\lambda_{\rm line} (1+z)^2 / H(z)$. All terms have implicit redshift dependence. A measurement of the mean surface brightness is equivalent to a luminosity density $\rho_{\rm line}(z) = 4 \pi \lambda_{\rm line} H(z) \bar S(z)$.

The SFR is observed to increase to $z \sim 2$ and decline by a factor of $\sim 10$ to the present \citep{2014ARA&A..52..415M}. In contrast, the dark matter structure in $\Lambda$CDM continues to grow over that period.  Neutral gas is the precursor to the molecular gas and evolves more gently than the SFR \citep[e.g.][]{2015MNRAS.452..217C}. To the extent it is understood, the average cold gas abundance also evolves differently from the SFR \citep[e.g.][]{2013ARA&A..51..105C}.

There is currently a wide range of predictions for mean CO brightness across cosmic time. See \citet{2016ApJ...817..169L} for a recent summary of models and assumptions. Simulations here use Model B of \citet{2013ApJ...768...15P} to define redshift evolution, down-scaled by two to approximately match COPSS-II \citep{2016ApJ...830...34K} observations. Calculations throughout assume $b_{\rm CO} =1.48$ \citep{2016ApJ...832..165C}. Predictions here scale the CO $J=1-0$ brightness by a factor of $10$ to be representative of the range of higher-$J$ transitions \citep{2013ARA&A..51..105C}. \citet{2016MNRAS.458L..99M} provides a model of monopole intensity from the full CO ladder. 

\citet{2014ApJ...793..116U} estimates \cii\ brightness through Eq.\,\ref{eqn:lirsb} using empirically determined relations \citep{2012ApJ...745..171S} for the line luminosity given the total IR luminosity, and the IR luminosity function of \citet{2011A&A...529A...4B}. They find that \cii\ surface brightness reaches a maximum of $\approx 5\kjysr$ by $z \approx 1$ and is a factor of $\approx 5$ lower by $z = 0$ and $z=2$. Following \citet{2014ApJ...793..116U}, we take a fiducial $b_{\rm CII} = 2$ \citep{2010A&A...518L..22C, 2012ApJ...750...37J}, on the low end of predictions \citep{2016ApJ...832..165C}. The interpretation of future data will require a complete model of the redshift evolution of the brightness and bias of $J=1-0$ and higher $J$ transitions of CO and the \cii\ line.

\subsection{PIXIE}

The approach here applies to general, deep surveys of the CMB spectrum. PIXIE provides a concrete example of parameters for next-generation instruments. PIXIE's primary scientific goals are to search for B-modes from inflationary gravitational waves, constrain large-scale E-modes, and measure spectral distortions of the CMB, such as the Sunyaev-Zel'dovich effect \citep{1969Ap&SS...4..301Z, 2015PhRvL.115z1301H, 2016SPIE.9904E..0WK}. To accomplish these goals, PIXIE will map the spectrum across the sky from $30$\,GHz to $6$\,THz using a symmetric, polarization-sensitive Fourier-transform Spectrometer (FTS) with heritage from FIRAS \citep{1994ApJ...420..439M}. PIXIE is forecast to have a factor of $\approx 1000$ greater sensitivity than FIRAS, driven mainly by (1) photon background-limited noise (through sub-Kelvin cooling), (2) controlled response to cosmic rays \citep{2016SPIE.9914E..1AN}, (3) larger etendue, and (4) increased sky and calibration integration time. PIXIE has four total detectors (two polarizations on each side of a symmetric FTS), but achieves high sensitivity through multimoded coupling (similar to FIRAS) to the FTS by light collectors.

Multimoded coupling results in sensitivity and instrumental simplicity at the expense of resolution. For our purposes, the beam is well-approximated by a \pixiefwhm\ Gaussian (see \citet{2014SPIE.9153E..18K} for the characterization of a beam model). Such large angular scales are compelling in cross-correlation because they trace linear perturbations where there is a clear connection to line brightness and bias.

\begin{figure}
\epsscale{1.2}
\plotone{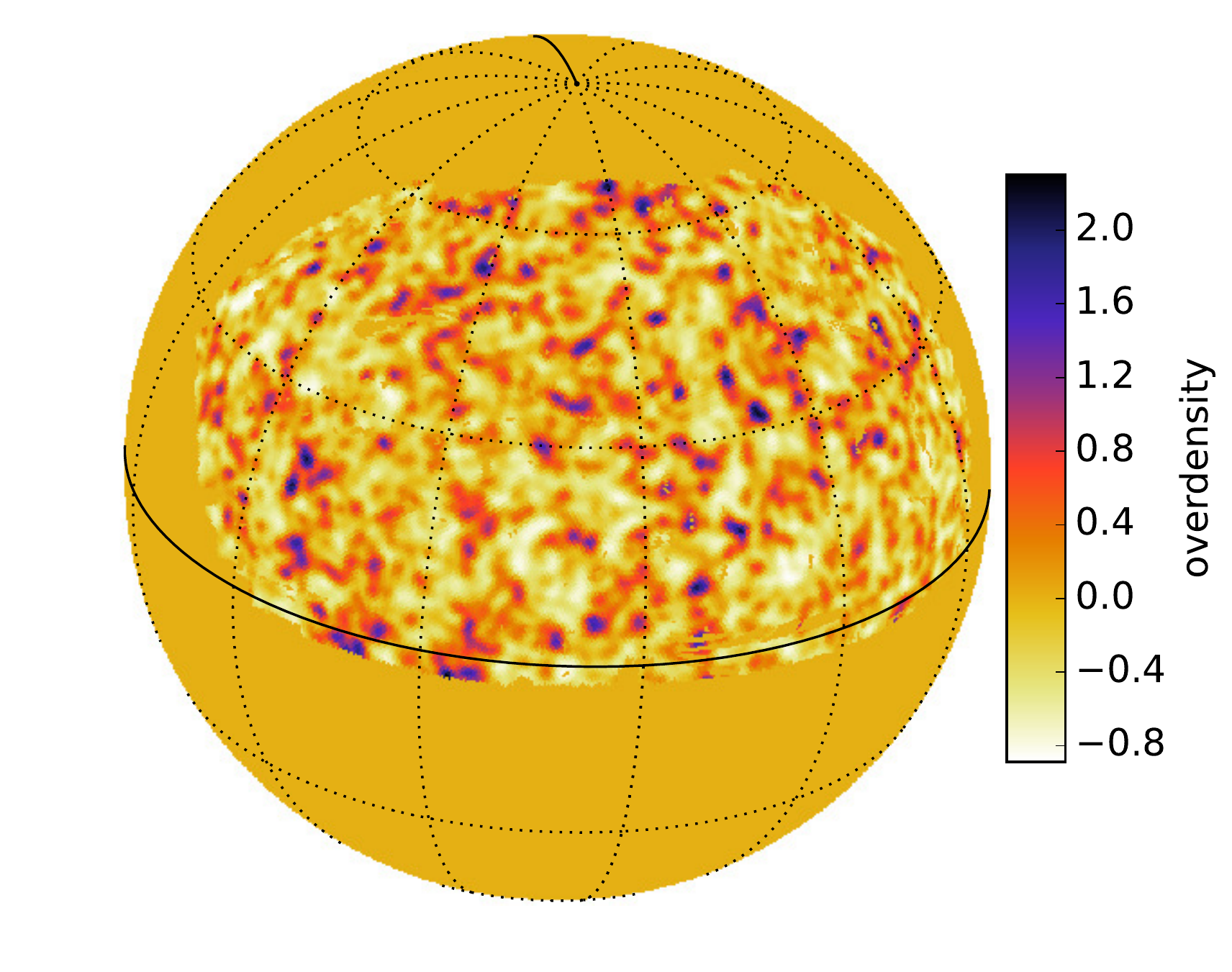}
\caption{The BOSS CMASS-North unitless overdensity $\delta$ in a slice of $0.51 < z < 0.53$, smoothed to PIXIE's \pixiefwhm\ effective beam, with graticules in celestial coordinates. The redshift range is equivalent to $\Delta \nu = 15$\,GHz for observations of \cii\ at $1245$\,GHz.}
\label{fig:delta}
\end{figure}

Fig.~\ref{fig:delta} shows the measured BOSS CMASS-North overdensity \citep[DR12][]{2015ApJS..219...12A} convolved by the PIXIE beam for a slice at $z\approx 0.5$, $\nu = 1245$\,GHz, and $\Delta \nu = 15$\,GHz. As an order-of-magnitude argument, multiplying the overdensity by $\sbcii \sim \kjysr$ \citep{2014ApJ...793..116U} gives intensity fluctuations similar to PIXIE's $\approx 1\kjysr$ noise per $1^\circ \times 1^\circ$ pixel. 

\begin{figure}
\epsscale{1.2}
\plotone{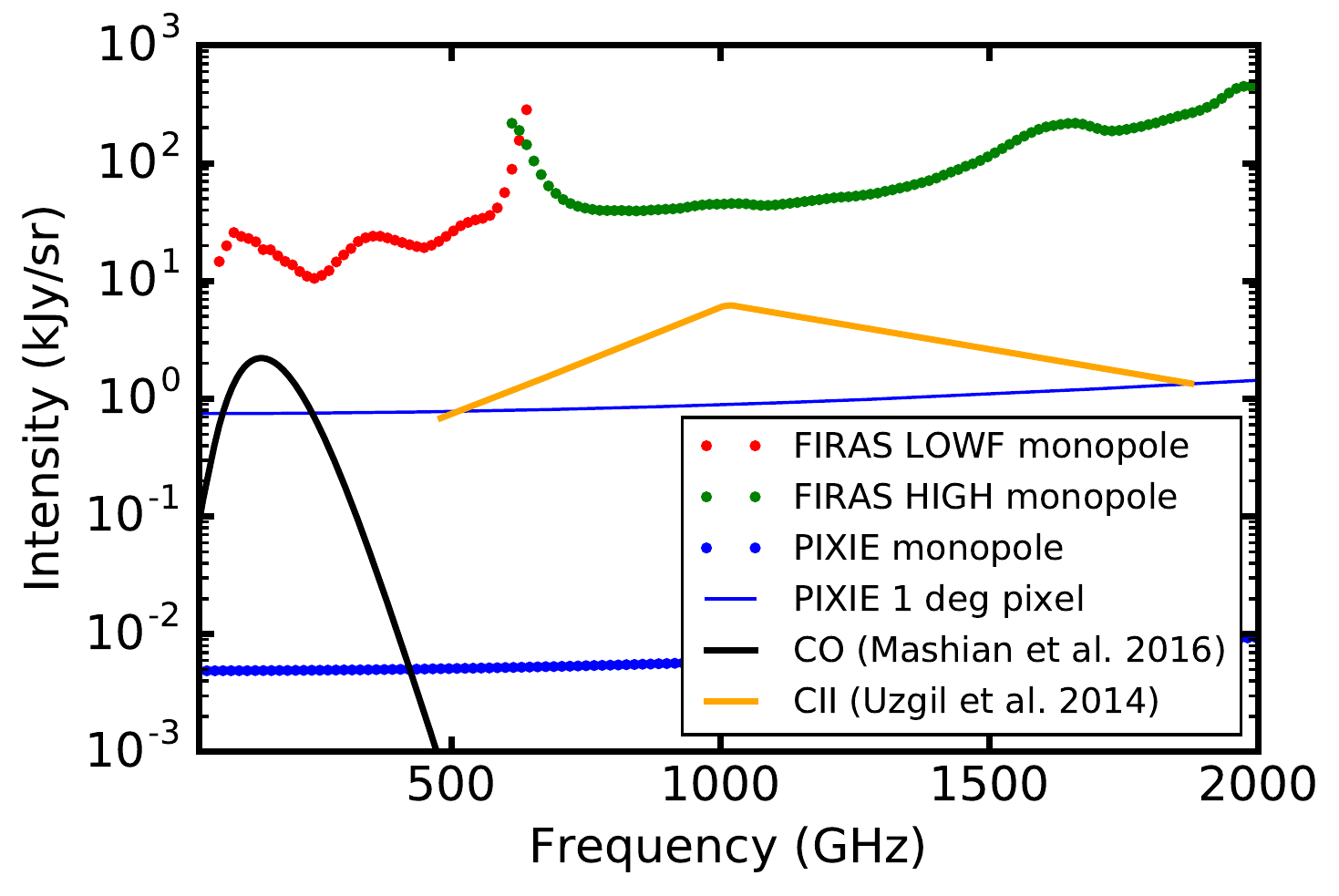}
\caption{Sensitivity of FIRAS (LOWF and HIGH) and PIXIE compared to predicted CO and \cii\ mean emission. The CO prediction \citep{2016MNRAS.458L..99M} is the cumulative spectral distortion over the ladder of lines. PIXIE's per-pixel sensitivity is comparable to the expected surface brightness of CO and \cii\ at mean density, and the monopole sensitivity is more than two orders of magnitude better.}
\label{fig:monopolesens}
\end{figure}

For completeness, note that CMB spectral distortion experiments have a unique sensitivity to the line emission monopole. \citet{2016MNRAS.458L..99M} first argued for PIXIE's sensitivity to global CO emission. The mean emission monopole is not a biased tracer of density, and a combination of monopole and anisotropy analysis could separate the brightness and bias. The bias of emitting gas is compelling on its own as an indicator of the underlying $M_{\rm gas}(M_{\rm halo})$ relation. Fig.~\ref{fig:monopolesens} shows PIXIE's monopole sensitivity and line brightness models from \citet{2016MNRAS.458L..99M} and \citet{2014ApJ...793..116U}. These are also broadly consistent with recent predictions by \cite{2016ApJ...833..153S}. The mean spectrum is not spatially modulated, making foreground separation more difficult. For these reasons, calculations here focus on the cross-correlation.

\section{Statistical constraints on the anisotropy}
\label{sec:statistical}

Intensity mapping literature has described constraints primarily on the 3D power spectrum, which exploit unique tomographic capabilities \citep[for 2D analysis, see e.g.][]{2013ApJ...768...15P, 2014MNRAS.443.3506B}. PIXIE's $15$\,GHz bands give fewer modes in $k_\parallel$ than $k_\perp$, and few total $k_\parallel$ modes for low-$J$ CO transitions. However, the primary scientific goal here is to determine the average line brightness as a function of redshift, rather than $P(|k|,z)$. This goal is better matched to 2D anisotropy analysis on each spatial slice at a constant frequency, which also simplifies the analysis. Predictions here use {\tt CLASSgal} \citep{2013JCAP...11..044D} to project the 3D matter power spectrum onto $C_\ell^{\delta \delta}$ in PIXIE's $15$\,GHz thick slabs. This algorithm directly integrates the projection at low $\ell$, where the Limber approximation is inaccurate.

Surface brightness fluctuations of redshifted line emission have three characteristic length scales. Scales $k \lesssim 0.1\, h{\rm Mpc}^{-1}$ track linear cosmological overdensity and correlations between, rather than within, halos \citep{2002PhR...372....1C}. On linear scales, surface brightness is a biased tracer of overdensity, as $\delta S = \sbgas \delta$, where $\bar S_{\rm line}$ is the mean surface brightness of the line and $b_{\rm line}$ is its bias. Following Eq.~\ref{eqn:lirsb}, $\sbgas$ constrains the first moment of the luminosity function.

For $k\gtrsim 0.1\, h{\rm Mpc}^{-1}$, correlations within a halo dominate and the power spectrum provides information about the line luminosity-weighted halo membership of galaxies. On smaller scales, shot noise of individual galaxies contributes a variance proportional to the inverse number density. The power spectrum on these shot-noise scales constrains the second moment of the luminosity function.  Halo-scale effects \citep{2016ApJ...817..169L} and shot noise \citep{2016ApJ...830...34K} provide additional degrees of freedom for estimating parameters in a line emission model.

Cross-correlation observations in the linear regime reduce the complications of nonlinear evolution and stochasticity between tracers (which must account for both one-halo effects and complex correlations of the shot noise between two galaxy populations). Signal in the linear regime directly measures $\sbgas$.

\subsection{Cosmic variance and the cross-power}

In surveys of the cosmic microwave background, a common strategy is to map each mode to ${\rm SNR}=1$ \citep{1997ApJ...480...72K}. Time is best spent integrating a larger number of modes rather than having high signal-to-noise on a mode that is ultimately limited by cosmic variance. Measurement of $\sbgas$ is more closely related to fitting for the amplitude of an overdensity template provided by the galaxy redshift survey. This determination of amplitude $\sbgas$ does not have cosmic variance. 

The harmonic two-point function conveniently accounts for the scale dependence of the beam, noise, and potentially stochasticity. The covariance of spherical harmonic modes of the galaxy and line intensity overdensity ($\delta^g_\ell$ and $\delta^{\rm IM}_\ell$) is 
\begin{equation}
{\mathbf \Sigma} = \left ( \begin{array}{cc} \alpha^2 C_\ell^{\delta \delta} + N^{\rm IM}_\ell & \alpha b_g C_\ell^{\delta \delta} \\ \alpha b_g C_\ell^{\delta \delta} & b_g^2 C_\ell^{\delta \delta} + N^{\rm shot}_\ell \end{array} \right ) \equiv \left ( \begin{array}{cc} C_\ell^{\rm IM} & C_\ell^{\times} \\ C_\ell^{\times} & C_\ell^{\rm gal} \end{array} \right ),
\label{eq:almcov}
\end{equation}
where $\alpha = \sbgas$, $C_\ell^{\delta \delta}$ is the matter overdensity power spectrum, $b_g$ is the bias of the galaxy redshift survey for $\delta$, and $N^{\rm shot}_\ell$ is the shot noise of the galaxy redshift survey. The second equivalence defines the auto-powers $C_\ell^{\rm IM}$ and $C_\ell^{\rm gal}$ of the intensity and galaxy surveys and the cross-power $C_\ell^{\times}$. All power spectra in Eq.\,\ref{eq:almcov} have been corrected for the CMB survey beam, which appears in $N^{\rm IM}_\ell = (\sigma^{\rm IM}_{\rm sr})^2 B_\ell^{-2}$, where $\sigma^{\rm IM}_{\rm sr}$ is the intensity map noise per steradian, and $B_\ell$ is the beam window function. The galaxy shot noise is the inverse of the number of galaxies per steradian, $(\bar n V_{\rm sr})^{-1}$, where $V_{\rm sr}$ is the volume of a $1$\,sr pixel with $\Delta \nu$ thickness and $\bar n$ is the counts density. 

On PIXIE's angular scales, current or future galaxy redshift surveys will have negligible shot noise to meet requirements for baryon acoustic oscillation measurements. For example, \pcb\ has the largest number of effective modes at $\ell \approx 60$, where the anisotropy of the measured BOSS overdensity is $\approx 10\times$ its shot noise. If not mentioned explicitly, shot noise will be neglected throughout for simplicity. Also, for simplicity, assume that uncertainty in $b_g$ is a negligible contribution to error in $\sbgas$ and is fixed to $b_g=2$ \citep{2011MNRAS.417.1350R, 2015MNRAS.451..539G}. In practice, a Bayesian approach should estimate all parameters in parallel, with the galaxy bias as a prior.

The Gaussian error on the cross-power is \cite[e.g.][]{2013ApJ...768...15P}
\begin{equation}
(\delta \hat C^\times_\ell)^2 = \frac{1}{M_\ell} [(C^\times_\ell)^2 + C_\ell^{\rm gal} C_\ell^{\rm IM}],
\label{eqn:fullcv}
\end{equation}
where $M_\ell \approx (2 \ell + 1) f_{\rm sky}$ is the number of modes per $\ell$. Finite survey area imposes some $\Delta \ell$ for bandpower binning, which multiplies $M_\ell$. Eq.\,\ref{eqn:fullcv} has been used to-date for predictions of intensity mapping cross-powers.

For studies of the average gas evolution, the quantity of interest is $\alpha = \sbgas$, not the full cross-power $C^\times_\ell = \alpha b_g C_\ell^{\delta \delta}$. Fitting the overdensity template $\delta$ to the intensity map determines $\sbgas$ without cosmic variance and has uncertainty per $\ell$
\begin{equation}
\sigma_\alpha^2(\ell) = \frac{1}{M_\ell} \frac{N_\ell^{\rm IM}}{C_\ell^{\delta \delta}}
\label{eqn:perlvar}
\end{equation}
and combined across all $\ell$,
\begin{equation}
\sigma_\alpha = \frac{\sigma^{\rm IM}_{\rm sr}}{\sqrt{M_{\rm tot}}} \frac{1}{\rm rms}
\label{eqn:sigma_a}
\end{equation}
with
\begin{eqnarray}
M_{\rm tot} &\equiv& \sum_\ell (2 \ell + 1) f_{\rm sky} B_\ell^{2} \nonumber \\
{\rm rms}^2 &\equiv&  M_{\rm tot}^{-1} \sum_\ell (2 \ell + 1) f_{\rm sky} B_\ell^{2} C_\ell^{\delta \delta},
\label{eqn:simple_sens}
\end{eqnarray}
where $M_{\rm tot}$ is the effective number of modes subject to resolution limits (the number of beam spots in the survey area), and ${\rm rms}$ is the effective rms per $\ell$-mode of the overdensity field. In combined CMASS North and South ($f_{\rm sky} \approx 0.25$), $M_{\rm tot} \approx 1670$. Eq.\,\ref{eqn:sigma_a} assumes white instrument noise, and the $\ell$-by-$\ell$ error Eq.\,\ref{eqn:perlvar} must be used for more general noise covariance such as residual foregrounds, described in Sec.\,\ref{sec:excessvar}. \citet{2011MNRAS.416.3009B} give a more generic Fisher matrix amplitude error in a scenario that includes the shot noise of the galaxy survey. 

\begin{figure}
\epsscale{1.2}
\plotone{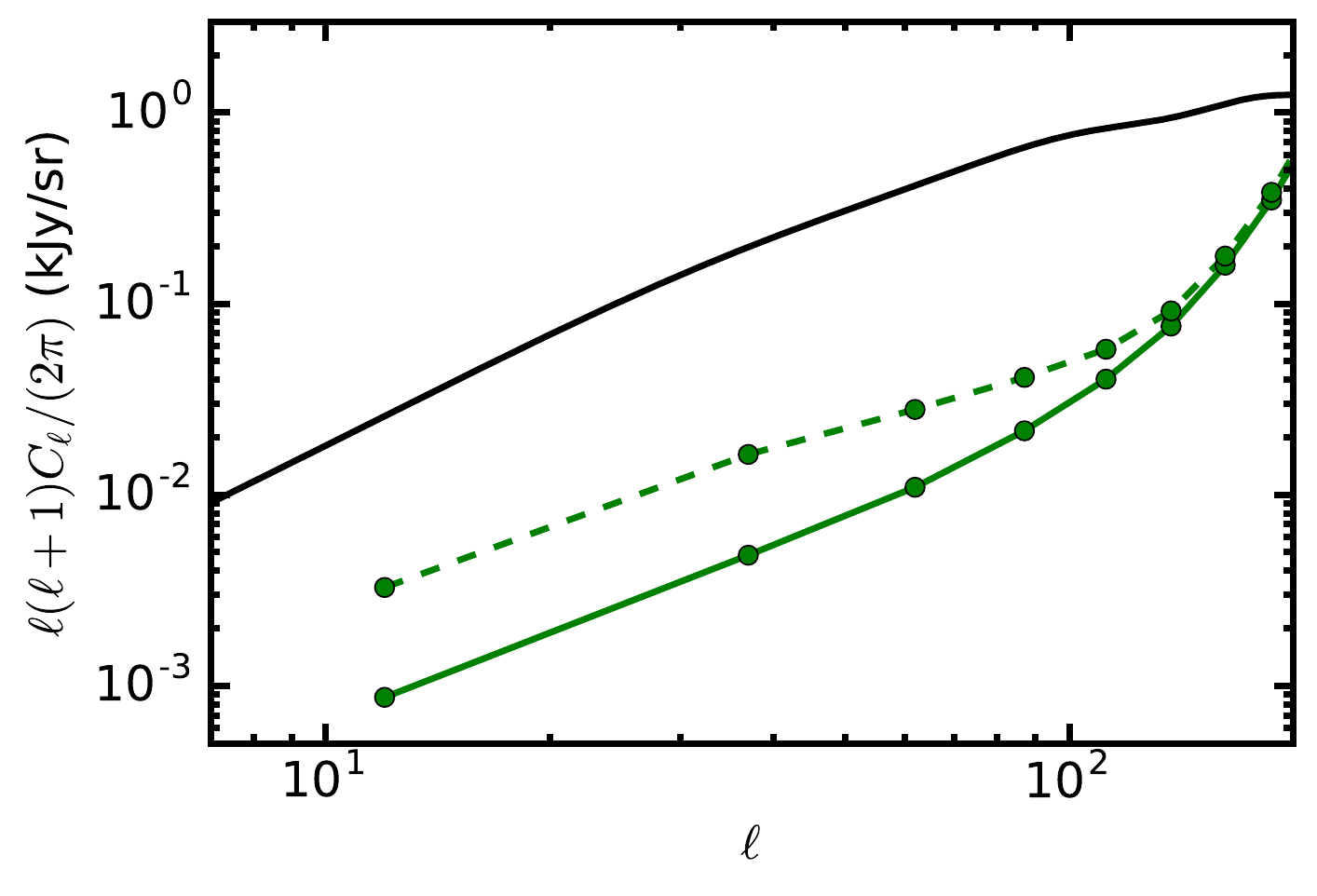}
\caption{Simulated \cii\ cross-power (black) of \pcb\ for the CMASS-North $z\approx 0.5$ slice in Fig.~\ref{fig:delta} with $\sbcii = 4\kjysr$ (taking $\bar S_{CII}=2\kjysr$ and $b_{CII}=2$). Green lines show errors without (solid) and with (dashed) cosmic variance in bins with $\Delta \ell =25$. The PIXIE beam window is corrected, causing the noise to increase beyond $\ell = 100$.
}
\label{fig:cv_errors}
\end{figure}

\subsection{Projected sensitivity}

Fig.~\ref{fig:cv_errors} shows the cross-power and errors for simulations of \pcb. The full cross-power error of Eq.~\ref{eqn:fullcv} is larger than Eq.~\ref{eqn:perlvar}, which measures $\sbgas$ without cosmic variance. The modeled cross-power and errors are consistent with power spectra \citep[estimated as][]{2002ApJ...567....2H} of simulations of BOSS and PIXIE data, which use the measured BOSS overdensity (Fig.~\ref{fig:delta}), simulated PIXIE maps (BOSS convolved by the PIXIE beam with PIXIE noise added), and BOSS random galaxy catalogs for shot noise. Simulations here use BOSS DR12 data \citep{2015ApJS..219...12A} only to provide a concrete example of density slices and mock cross-correlation.

\begin{figure}
\epsscale{1.2}
\plotone{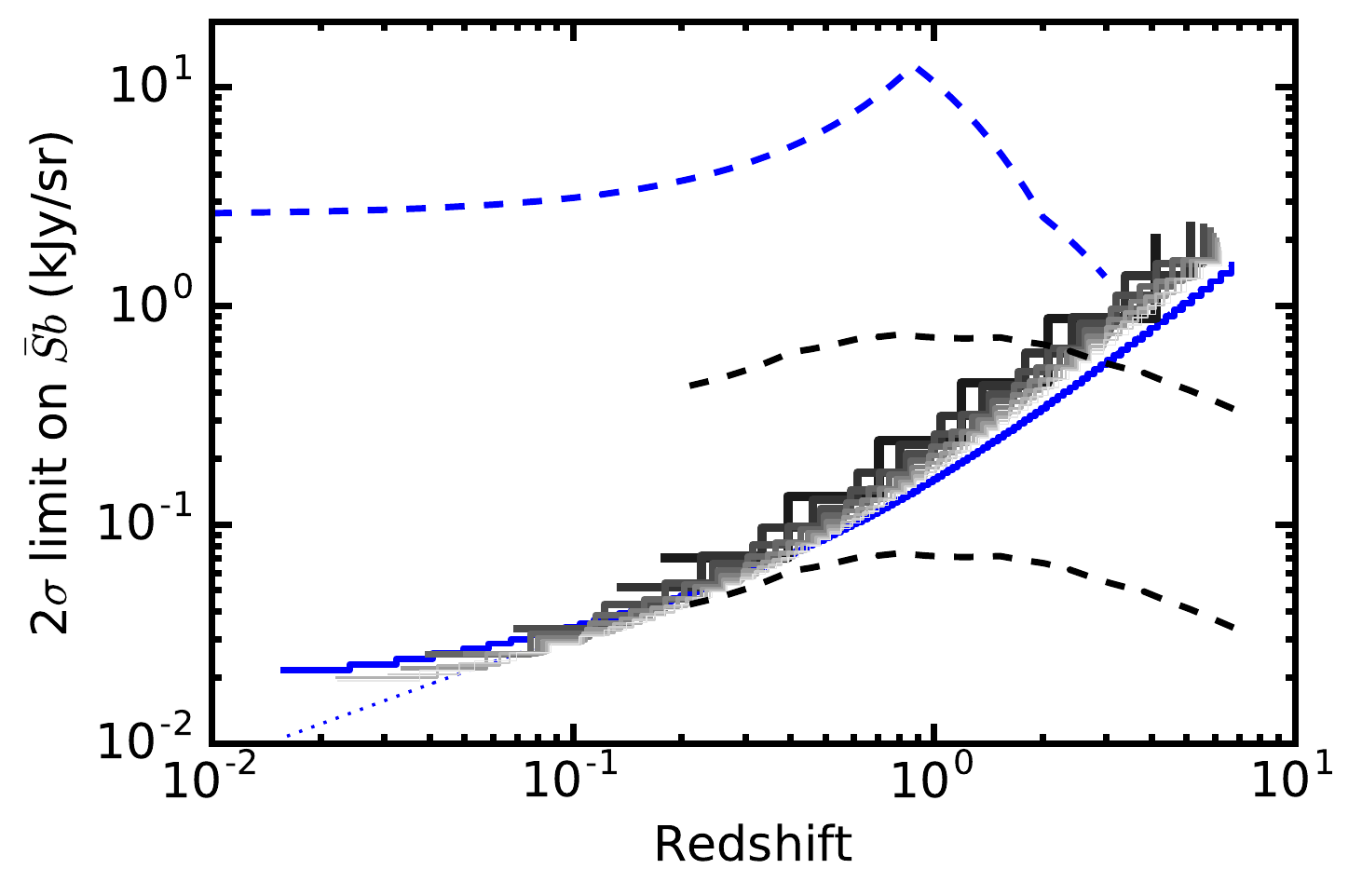}
\caption{PIXIE $2\sigma$ sensitivity to CO (black for $J=1-0$ to light gray moving up the ladder) and \cii\ (blue) for individual spectral bins. $f_{\rm sky} = 0.25$. The CO constraint improves toward higher $J$ as the $\Delta \nu / \nu_{\rm line}$ ($\Delta \nu = 15$\,GHz) slab thickness decreases. The CO $J=1-0$ model (lower black dashed line) is model B from \citet{2013ApJ...768...15P} divided by two to agree with COPSS\,II \citep{2016ApJ...830...34K} measurements, and multiplied by $b_{\rm CO} = 1.48$ \citep{2016ApJ...832..165C}. The CO SLED is more luminous toward higher-$J$ transitions. The upper black line represents a $10\times$ multiplier for the $J=5-4$ transition relative to $J=1-0$ typical of sub-millimeter galaxies \citep{2013ARA&A..51..105C}. The \cii\ (blue dashed) brightness model is from \citet{2014ApJ...793..116U} multiplied by $b_{\rm CII} = 2$. The blue dotted line shows \cii\ sensitivity including nonlinear structure.}
\label{fig:sensitivity_summary}
\end{figure}

Fig.~\ref{fig:sensitivity_summary} compares sensitivity from Eq.~\ref{eqn:simple_sens} to expected brightness in the CO ladder and \cii. This assumes $f_{\rm sky}=0.25$, sensitivity limited by instrumental noise, and that shot noise in the galaxy redshift survey is negligible on these large scales. Hence these predictions are fairly generic with regard to the galaxy survey in cross-correlation, scaling as $1/\sqrt{f_{\rm sky}}$. Future wide-area spectroscopic surveys \citep{2014JCAP...05..023F} and photometric surveys (with $\sigma_z \ll \Delta z$ bins) could support cross-correlation to $z\sim 3$. This redshift range covers the rise and fall of the SFR. PIXIE's resolution and requisite large survey area are not well-matched to studies of reionization. 

At higher redshifts, the rms of the overdensity field diminishes because there has been less growth of structure, and that structure is at smaller angular scales, which are impacted by the beam. Higher $J$ transitions have improved constraints through lower $\Delta \nu / \nu_{\rm line}$ and brighter expected intensities. CO sensitivity can surpass \cii\ at low redshift (despite being thicker slabs) because PIXIE noise increases at high frequency (Fig.~\ref{fig:monopolesens}). Sec.\,\ref{sec:excessvar} describes contamination and the impact of residuals after cleaning. 

Including nonlinear contributions through {\tt halofit} \citep{2003MNRAS.341.1311S} improves the constraints at $z=0.05$ by $\approx 22\%$, as shown in Fig.~\ref{fig:sensitivity_summary}. At these low redshifts, the PIXIE \pixiefwhm\ beam can subtend nearby nonlinear scales. Predictions here neglect both nonlinear evolution and line emission shot-noise contributions. Additional variance from these effects improves the predicted signal sensitivity, making predictions here conservative. $\sigma^{\rm IM}_{\rm sr}$ for PIXIE used here is the average across the sky, but the scan strategy has additional depth at the ecliptic poles. 

\subsection{Scale-dependence}

Intensity fluctuations do not track the matter density perfectly on all scales. Including this as stochasticity $r_\ell = C_\ell^\times / \sqrt{C_\ell^{\rm IM} C_\ell^{gal}}$ (where $C_\ell^{\rm IM}$ and $C_\ell^{gal}$ are the intensity and galaxy auto-powers without shot or instrument noise), the intensity cross-correlation constrains $\alpha_\ell  = \sbrgas$ at a given $\ell$. On the linear scales probed by PIXIE, line emission is expected to be a biased tracer of the same overdensities probed by the galaxies, so that $r_\ell \approx 1$ \citep{2016MNRAS.458.3399W}. The stochasticity departs from $1$ on one-halo and shot scales due to differences in halo occupation. 

A convenient, $\ell$-by-$\ell$ estimator for $\alpha_\ell = \sbrgas$ is 
\begin{equation}
\hat \alpha_\ell = \frac{\hat C^\times_\ell}{(\hat C_\ell^{gal} - N_\ell^{\rm shot}) b_g^{-1}},
\label{eq:lbyl}
\end{equation}
where $\hat C_\ell^{gal}$ is the measured power spectrum of the galaxy survey (from which $N^{\rm shot}_\ell$ is removed), and $\hat C^\times_\ell$ has been corrected for the CMB survey beam. The denominator has one factor of the galaxy bias that cancels with the numerator, $\langle \hat C^\times_\ell \rangle\,=\,\sbrgas b_g C_\ell^{\delta \delta}$. The numerator and denominator scatter in the same direction due to cosmic variance. Plots of $\hat \alpha_\ell$ provide a diagnostic for possible stochasticity toward small scales. This form of $\hat \alpha_\ell$ has the advantage of using the measured density fluctuations in the slice to determine the line brightness amplitude, rather than a model for $C_\ell^{\delta \delta}$. Inference of $\sbgas$ from the auto-power alone requires an accurate model of the power spectrum.

\section{Contamination}
\label{sec:excessvar}

Forecasts in Sec.~\ref{sec:statistical} account for random instrumental noise and show optimal sensitivity limits. The voxels in the spectral survey contain surface brightness from all other sources of emission, which can either degrade the sensitivity or produce bias. For example, the FIRAS average surface brightness in the CMASS-North region is $3.2\mjysr$ at 1270\,GHz ($z = 0.5$ for \cii\ emission near the peak of CMASS $\bar n(z)$, shown in Fig.\,\ref{fig:delta}). The required magnitude of contamination removal is similar to $21$\,cm tomography. Calculations below describe contamination at $\nu > 600$\,GHz and $\nu < 600$\,GHz, which are relevant for \cii\ and CO, respectively, and have qualitatively different contributions. Given the sensitivity margins to \cii\ shown in Fig.~\ref{fig:sensitivity_summary}, $\nu > 600$\,GHz is described in more detail. Galactic emission is significantly brighter than the line emission and limits the region of the sky, hence the largest accessible angular scale. Instrument response to bright contamination can also result in unmodeled residuals after cleaning. Sec.\,\ref{ss:instfg} describes calibration and beam requirements to control these residuals.

The line intensity signal is uncorrelated with both the Galaxy and most of the variance in the extragalactic anisotropy. Residuals after cleaning therefore increase the error bars but do not bias the cross-correlation with a galaxy redshift survey. The goal of cleaning becomes one primarily of removing variance from the maps. Continuum contamination is well-described by a limited set of smooth spectral functions, while the signal can vary from channel-to-channel. Simulations here show that simple template subtraction and linear combinations of channels remove much of the contamination. An alternative approach fits the contamination spectrum along each line of sight, either parametrically \citep[e.g.][]{2008ApJ...676...10E} or blindly \citep[e.g.][]{2015ApJ...815...51S, 2017MNRAS.464.4938W}. Sec.\,\ref{ss:corrcib} describes continuum contamination that is correlated with line signal. This class is removed well by continuum cleaning and can be characterized by correlations of spatial slices at offsets in frequency.

Unlike the cross-power, the auto-power is additively biased by uncorrelated residuals from cleaning. A primary challenge of autonomous intensity mapping surveys is in ruling out this bias. The cross-power is formally a lower bound on $\sbgas$ because $r_\ell \leq 1$. A lower bound from the cross-power and an upper bound from the auto-power sandwich the true line brightness \citep{2013MNRAS.434L..46S}.

\subsection{Galactic contamination $\nu > 600$\, GHz}
\label{ss:galcont}

\begin{figure}
\epsscale{1.2}
\plotone{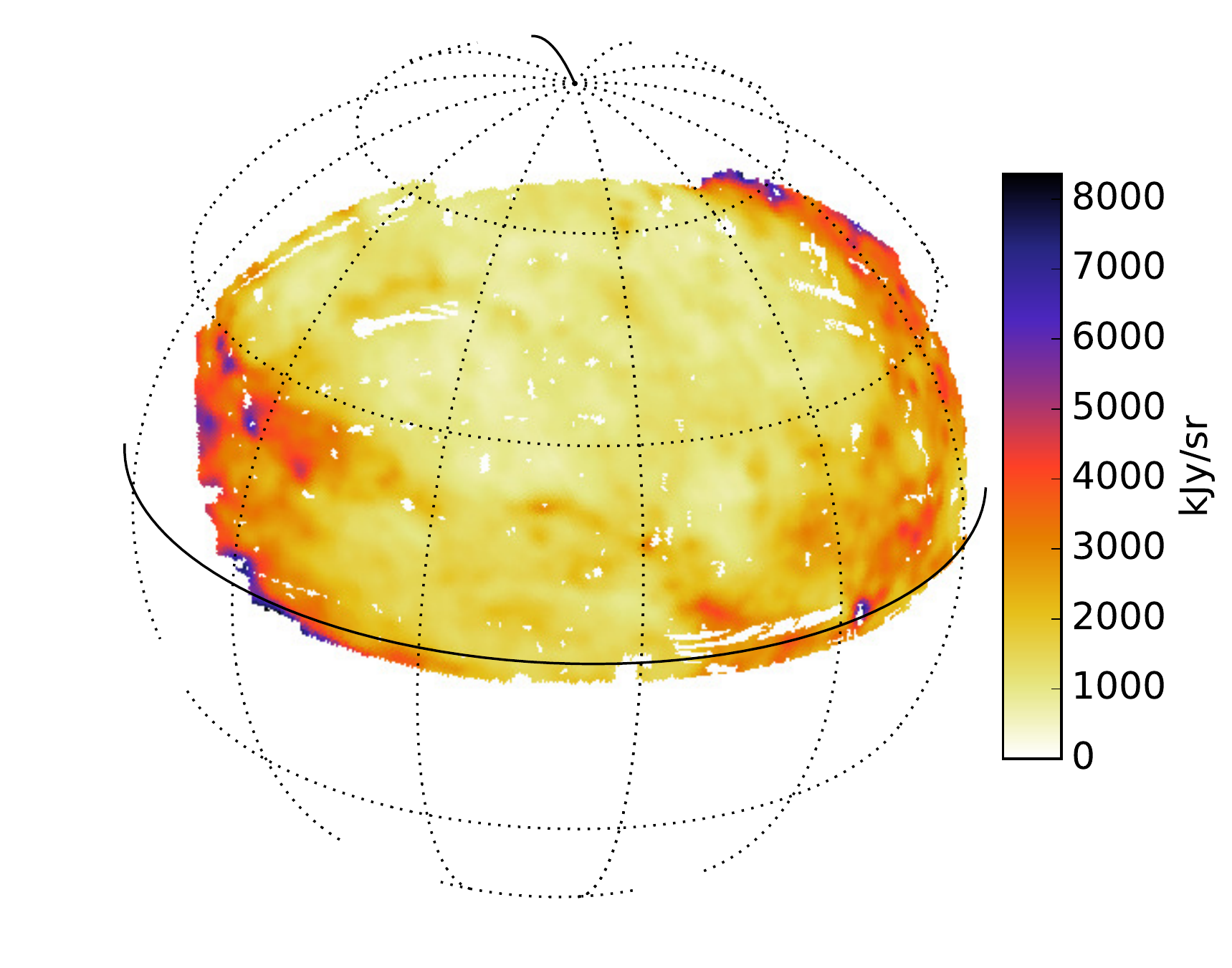}
\caption{Thermal dust emission from the galaxy at $1245$\,GHz (\cii\ at $z \approx 0.5$) in the same BOSS CMASS-North region as Fig.\,\ref{fig:delta}, from the model of \citet{2016arXiv160802841T}, convolved by the PIXIE beam and with PIXIE noise added.
}
\label{fig:galaxy}
\end{figure}

Galactic and extragalactic thermal dust emission dominate contamination at $\nu > 600$\,GHz. Simulations here begin with only the galactic contribution and add progressively more components to isolate their behavior. Fig. \ref{fig:galaxy} shows the single-component dust model from {\tt PySM} \citep{2016arXiv160802841T} at $1245$\,GHz, convolved to the PIXIE beam. This model assumes $I = A\nu^{\beta_d} B_\nu(T_d)$ with Planck function $B_\nu$, where the amplitude $A$, index $\beta_d$, and temperature $T_d$ vary spatially \citep{2015A&A...576A.107P}. $1245$\,GHz is the reference frequency here because it lies near the maximum of galactic and extragalactic dust surface brightness, and the maximum $\bar n(z)$ for \cii\ in BOSS DR12 data, which provide an example survey and overdensity.

Model the slice in a reference frequency $\nu$ as ${\bf s}_\nu = {\bf V} {\bf a}_\nu + {\bf n}_\nu$, where ${\bf s}_\nu$ is an $N_{\rm pix}$-long spatial map vector at $\nu$, ${\bf V}$ $(N_{\rm pix} \times N_{\rm veto})$ is a set of maps of contamination, which have amplitudes ${\bf a}_\nu$  ($N_{\rm veto}$ values), and the noise in the slice is ${\bf n}_\nu$ $(N_{\rm pix})$. The linear combination amplitudes which minimize residual variance are $\hat {\bf a}_\nu = ({\bf V}^T {\bf N}^{-1} {\bf V})^{-1} {\bf V}^T {\bf N}^{-1} {\bf s}_\nu$ for covariance ${\bf N} = \langle {\bf n}_\nu {\bf n}_\nu^T \rangle$. The channels that clean the reference band in a linear combination will be referred to as ``veto" channels to emphasize that they are not necessarily contamination component templates. Let ${\mathbf \Pi} = {\bf V} ({\bf V}^T {\bf N}^{-1} {\bf V})^{-1} {\bf V}^T {\bf N}^{-1}$ project onto the veto channels. A map ${\bf s}_\nu$ can then be cleaned with the linear combination as ${\bf s}_\nu^{\rm clean} = {\bf s}_\nu - {\bf V} \hat {\bf a}_\nu = (1-{\mathbf \Pi}) {\bf s}_\nu$.  This simple cleaning approach shows the magnitude of cleaning considerations. At the level of this demonstration, the fidelity of contamination and instrument models are a greater limitation than the cleaning approach.

Find the impact of residual contamination after cleaning by (1) linearly estimating amplitudes of the veto maps in the reference science band (assuming diagonal noise covariance), (2) subtracting ${\bf V} \hat {\bf a}_\nu$, (3) calculating cross- and auto-power spectra in the masked region with MASTER \citep{2002ApJ...567....2H}, and (4) using Eq.\,\ref{eqn:perlvar} to find the increase in the error on $\sbcii$ ($\sigma_{\bar S b, {\rm CII}}$) due to additional variance from residuals in excess of instrumental noise. In the more general case with residuals, the $N_\ell^{\rm IM}$ in Eq.\,\ref{eqn:perlvar} is replaced by the estimated auto-power after cleaning (signal-free).
 
Using the linear combination of the $1185$\,GHz channel to clean $1245$\,GHz, residual dust emission from the galaxy in the CMASS-North region ($f_{\rm sky} = 0.18$) results in $2 \sigma_{\bar S b, {\rm CII}} = 2.8\kjysr$, a factor of $25$ greater than $2 \sigma_{\bar S b, {\rm CII}}$ from instrumental noise alone. (Separating $1185$\,GHz and $1245$\,GHz by four channels reduces signal correlations with the veto band, described in Sec.\,\ref{ss:sigtemp} and Fig.\,\ref{fig:corrcib}). Adding the $1305$\,GHz channel to the linear combination results in $2 \sigma_{\bar S b, {\rm CII}} = 0.14\kjysr$, or only a $25\%$ increase over instrumental noise. Most of this increase in noise is a result of the uncorrelated thermal noise in the linear combination of bands. Residual galactic contamination in this two veto-channel cleaning contributes $5\%$ over the instrumental noise. A second map in the linear combination provides a degree of freedom that explains residuals produced by spatial variation of the emissivity. The thermal dust model of \citet{1999ApJ...524..867F} produces a similarly small degradation in $\sigma_{\bar S b, {\rm CII}}$ in the two-band subtraction approach, again because a spatially varying amplitude and index describe most of the variance in channels near $1245$\,GHz.

\begin{figure}
\epsscale{1.2}
\plotone{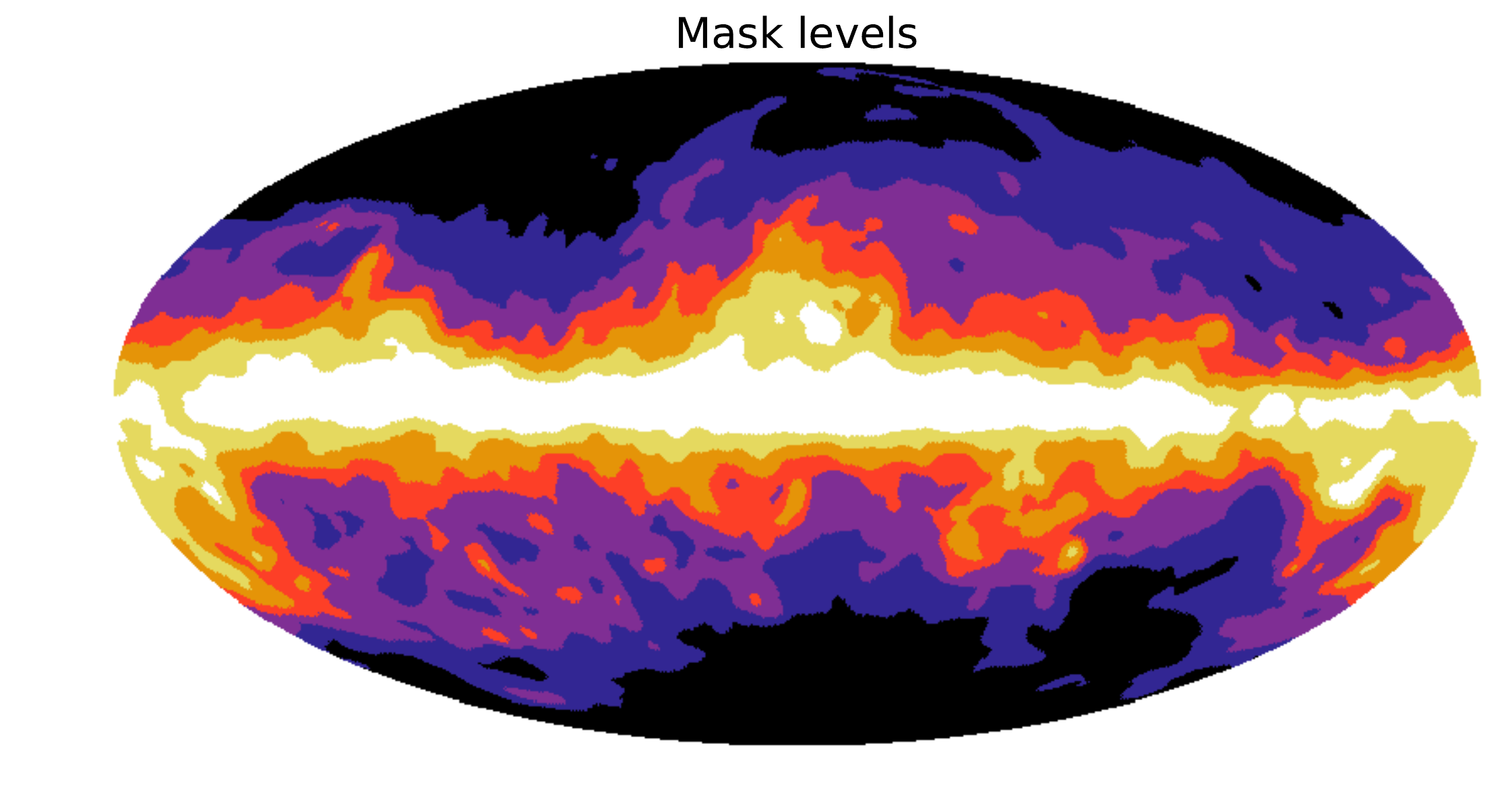}
\epsscale{1.2}
\plotone{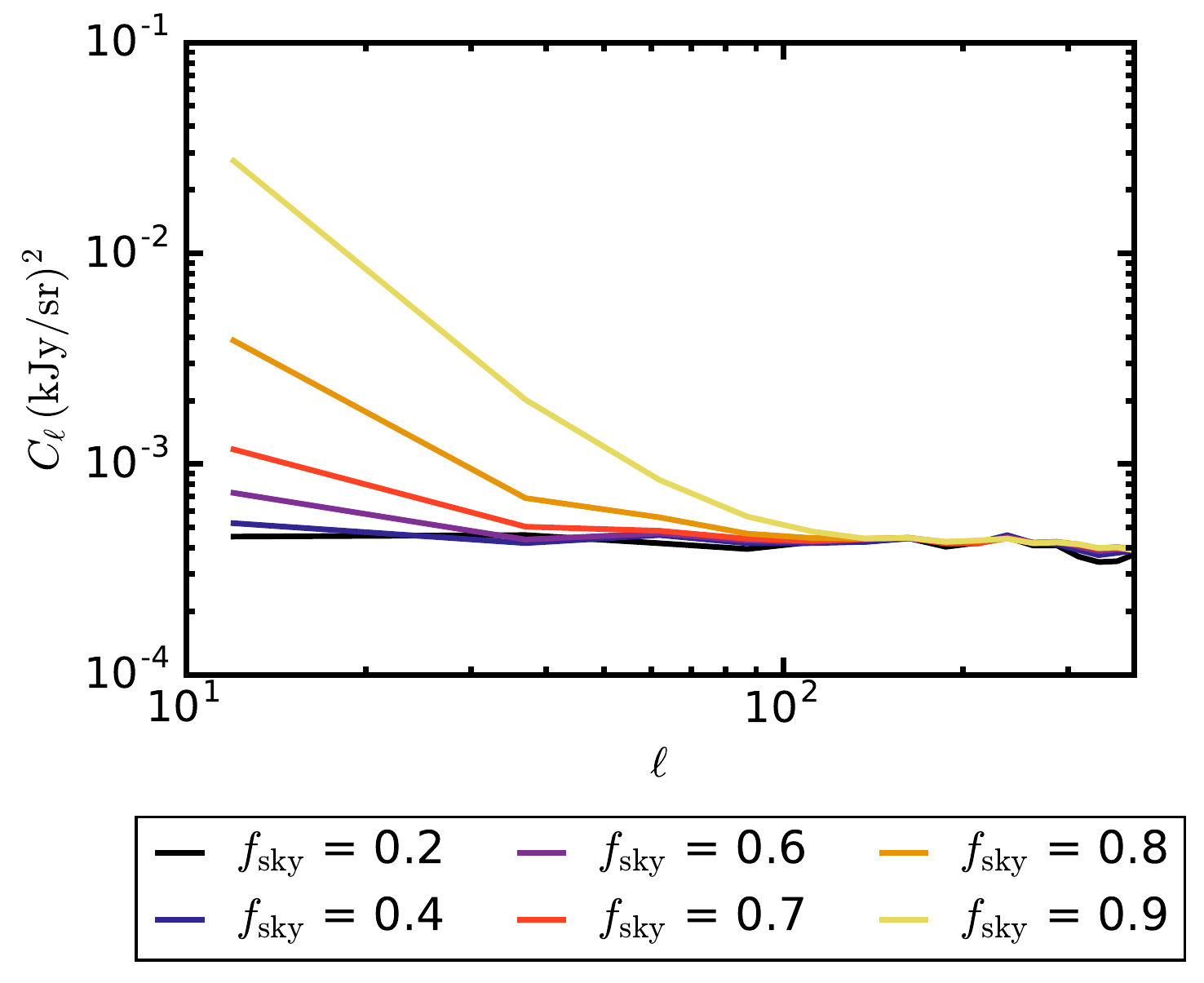}
\caption{Upper: regions corresponding to $f_{\rm sky} = \{ 0.2, 0.4, 0.6, 0.7, 0.8, 0.9 \}$ of the cleanest regions of the sky, starting from black for $f_{\rm sky} = 0.2$ and adding regions in lighter colors. Lower: auto-power spectrum of residual galactic contamination at $1245$\,GHz after cleaning with a linear combination of $1185$ and $1305$\,GHz. Residuals from the Galaxy increase errors in the cross-power but are uncorrelated with the cosmological signal, and thus do not produce bias. On greater than the cleanest $70\%$ of the sky, galactic residuals add significant variance on large scales. $\Delta \ell = 25$ binning is chosen to be compatible with the smallest survey region. The spectrum does not undo beam convolution, so it reaches a plateau of instrumental noise variance at high $\ell$.}
\label{fig:galactic_resid}
\end{figure}

The two-band cleaning approach can be applied to progressively larger $f_{\rm sky}$ to test when the simple model here fails. Fig.~\ref{fig:galactic_resid} shows the sky divided into the patches of $f_{\rm sky} = \{ 0.2, 0.4, 0.6, 0.7, 0.8, 0.9 \}$ of the cleanest sky regions (with masks based on $3^\circ$ FWHM smoothing of the galaxy model). The lower panel shows the auto-power spectra for these masks after two-channel linear combination cleaning in the dust model $I = A\nu^{\beta_d} B_\nu(T_d)$. Galactic contamination adds variance on the largest angular scales for mask regions that exceed the cleanest $\sim 70\%$ of the sky. This additive residual variance in the auto-power produces larger errors in the cross-power with the galaxy survey but not bias. Fig.\,\ref{fig:instresp} shows that sensitivity to $\sbcii$ is modestly degraded because much of the weight comes from $\ell \approx 60$ rather than lower $\ell$.

The cleaning operation $(1-{\mathbf \Pi}) ({\bf s}_\nu^{\rm signal} + {\bf s}_\nu^{\rm fg})$ applies to both signal ${\bf s}_\nu^{\rm signal}$ and foreground ${\bf s}_\nu^{\rm fg}$, so the cleaned map contains non-zero $-{\mathbf \Pi} {\bf s}_\nu^{\rm signal}$. This quantity is anticorrelated with signal due to spurious correlation of overdensity and the veto channels. A further advantage of the cross-correlation approach is that the galaxy survey provides a map of overdensity signal as a proxy for ${\bf s}_\nu^{\rm signal}$ which can be used to estimate this bias, or $\approx 0.2 \sigma$ for the cases simulated here.

\subsection{Line intensity signal in the linear combination}
\label{ss:sigtemp}

The veto channels also contain cosmological line signal of approximately the same amplitude as the central science channel. The cosmological signal can be partly coherent between these slices due to large-scale structure at low $k_\parallel$, potentially causing the linear combination to project out some signal. The channels at $1185$ and $1305$\,GHz have negligible correlation with the central band (Fig.~\ref{fig:corrcib}). The uncorrelated signal in the veto adds variance, which increases $\sigma_{\bar S b, {\rm CII}}$ by a factor of $1.8$. Jointly model the signal and contamination by adding overdensity ${\boldsymbol \delta}$ derived from the galaxy redshift survey to the stack of maps, as ${\bf V} = \{ {\bf S}(\nu_l), {\bf S}(\nu_l), {\boldsymbol \delta}(\nu_l), {\boldsymbol \delta}(\nu_h), {\boldsymbol \delta}(\nu_o) \}$, where ${\bf S}$ are the intensity survey maps and $\nu_l = 1185$\,GHz, $\nu_o = 1245$\,GHz (reference), and $\nu_h = 1305$\,GHz. Using this choice of veto maps recovers the estimate of $\sbcii$ to within $1\%$ of the instrumental noise limit without statistically significant bias from signal correlations along $k_\parallel$. Note that the template ${\boldsymbol \delta}(\nu_o)$ fits the signal and must be added back. (Alternately, in the context of a joint likelihood on signal and contamination, the amplitude of ${\boldsymbol \delta}(\nu_o)$ constitutes an estimate of $\sbcii$.) This calculation assumes that the galaxy overdensity is a perfect proxy for the line intensity signal, but in practice $\sigma_{\bar S b, {\rm CII}}$ may be degraded by stochasticity $r_\ell$.

Interlopers from other redshifts will increase errors but not bias the cross-power with a galaxy redshift survey. If a galaxy redshift survey can provide slices of overdensity at redshifts of known interlopers, these could be added as templates to reduce variance.

\subsection{Interaction with the instrument}
 \label{ss:instfg}

Spectral response calibration and stability are essential in intensity mapping experiments and must be controlled at the level of signal-to-contamination. For example, a $1\kjysr$ fluctuation could be either a fluctuation in $\sbcii \delta$, or a $0.1\%$ fluctuation in the response to $1\mjysr$ contamination. Cross-correlation can still extract a signal that is coherent with the galaxy overdensity and remain unbiased, but residual contamination decreases sensitivity in the cross-power.

Precision calibration is already an objective for the primary science goal of CMB spectral distortions. The differential FTS architecture for CMB spectral distortions measures surface brightness relative to a calibration source, which can have a well-characterized, stable, and smooth spectral shape. The calibrator can be integrated as deeply as the sky. The proposed PIXIE calibrator is black to $3 \times 10^{-7}$ \citep{2014SPIE.9143E..1EK}, driven by multiple bounces in a conical light trap similar to FIRAS. A constant, absolute temperature error or spectral structure of the calibrator will induce spectral gain errors that are spatially constant. Cleaning demonstrated here is robust to monopole spectral gain errors, which are solved in the veto-channel amplitudes. The most challenging systematics have both spatial and spectral structure. Instability in the instrument or calibration becomes a spatial structure in the maps. \citet{1994ApJ...420..457F} and \citet{FIRASEXSUPP} describe a general set of systematic terms in FIRAS. The differential nature of the FTS and operation near the CMB temperature both suppress the impact of instrumental emission. FIRAS analysis solved for emission terms of the instrument frequency-by-frequency at the level of $\approx 0.2\%$ in gain \citep{FIRASEXSUPP}. Detector model errors can also produce gain errors with relatively smooth spectral structure. Variation in system temperatures can then produce spatial modulation. Detector model errors are $0.4\%$ in the CMB bands and reach $2\%$ toward $2$\,THz.

\begin{figure}
\epsscale{1.2}
\plotone{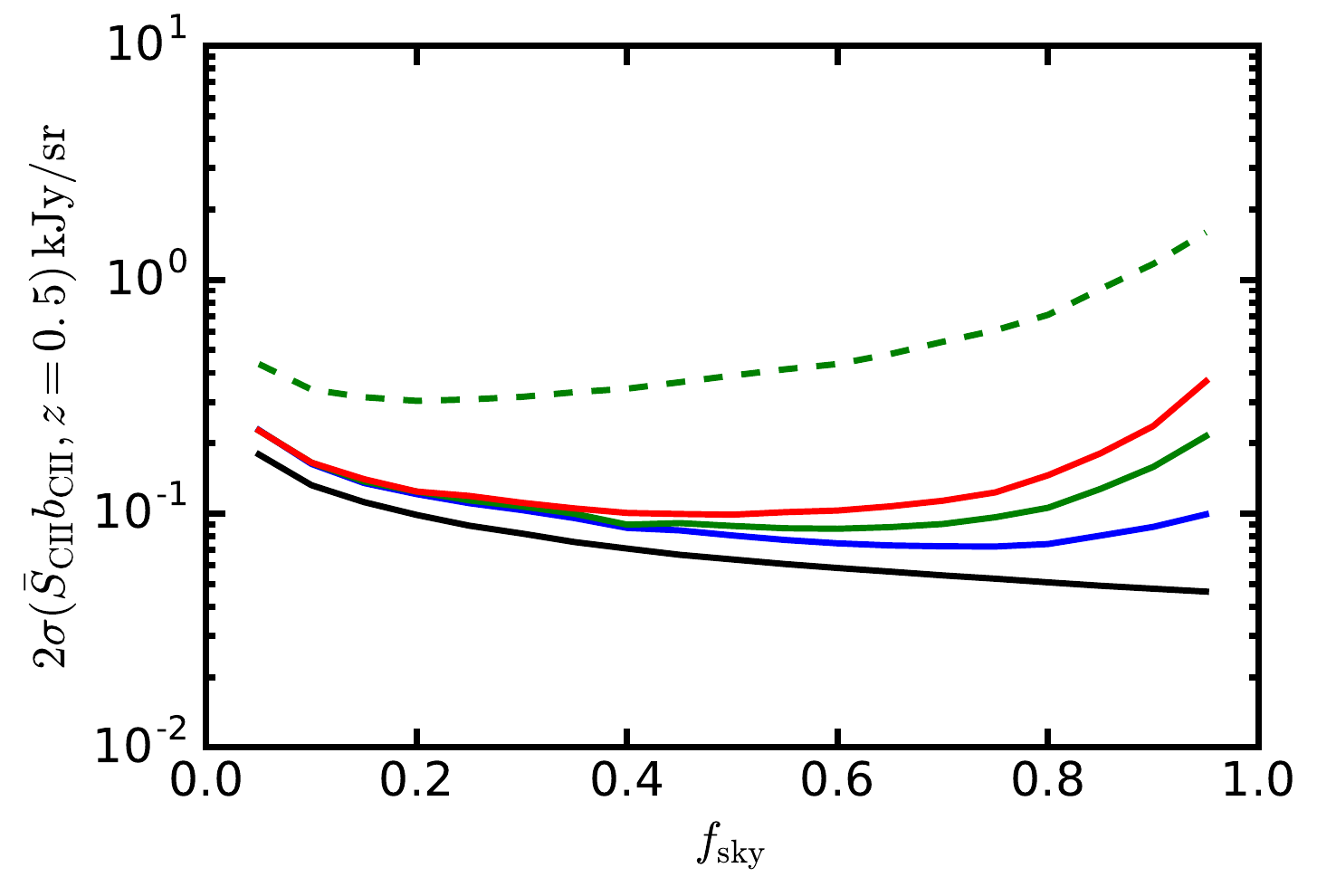}
\caption{$2\sigma$ constraint on $\sbcii$ at $z=0.5$ and the influence of instrument response, as a function of sky area. The black curve shows the limit from instrumental noise, and blue adds galactic foregrounds and cleaning. The two-channel subtraction from Sec.~\ref{ss:galcont} cleans the majority of foreground emission. The upper green dashed curve shows the impact of $4 \times 10^{-3}$ variations in gain calibration that are spatially and spectrally uncorrelated. The lower green curve reduces this uncorrelated gain model error to $4 \times 10^{-4}$. The red curve shows the impact of $0.5\%$ measurement error or variation in beam solid angle between bands. Instrumental model residuals tend to produce more contamination in bright regions of the galaxy and so limit the maximum $f_{\rm sky}$.}
\label{fig:instresp}
\end{figure}

Fig.~\ref{fig:instresp} shows the impact of gain variations at $4\times 10^{-3}$ and $4\times 10^{-4}$, in a worst-case scenario of independent gain fluctuations in each spatial/spectral pixel. Under the assumptions here, control to $4\times 10^{-4}$ is sufficient to add modest noise to the cross-power. Detector noise and external calibrator signal integration limited the FIRAS calibration. PIXIE calibration at THz frequencies is expected to be better than $10^{-5}$ with no sharp (band-to-band) features.\footnote{A. Kogut 2017, private communication.} A single calibration cone heated to $20$\,K provides additional calibration signal in the Wien tail of a calibrator, which is otherwise at the CMB temperature \citep{2016SPIE.9904E..0WK}.

Chromatic variation of the beam can mix spatial into spectral structure, and partly destroy the coherence of continuum contamination across frequency channels, which is essential to the cleaning pursued here. An approach to chromaticity in fully sampled images is to convolve the maps to the lowest, common resolution based on a model of the beam (used in \citet{2013MNRAS.434L..46S} for GBT data). Fig.~\ref{fig:instresp} shows that the solid angle must be compensated between bands to $<0.5\%$ tolerance (similar to that achieved in Planck HFI \citet{2014A&A...571A...7P}) to suppress inter-band residuals to a level that contribute less than $\sigma_{\bar S b, {\rm CII}}$ from instrumental noise. For FTS designs such as PIXIE and FIRAS, the beam shape is dominated by the concentrator and approximately convolved by the diffraction scale \citep{2014SPIE.9153E..18K}. At $>600$\,GHz, the beam is significantly less chromatic than the diffraction-limit.

\subsection{Extragalactic Contamination $\nu > 600$\, GHz}

Simulate CIB emission as a realization of the power spectrum of the form \citep{2017MNRAS.466..286M}
\begin{equation}
\frac{\ell (\ell +1)}{2 \pi} C^{\rm CIB, Mak}_\ell \propto \left ( \frac{\ell}{2000} \right)^{0.56},
\label{eqn:makcib}
\end{equation}
extrapolated to frequencies of interest using the graybody law from \citet{2012ApJ...752..120A}. The overall Spectral Energy Distribution (SED) scaling is taken to be the best-fit to amplitudes inferred from Planck data \citep{2017MNRAS.466..286M} at $353$, $545$, and $857$\,GHz. In a reference band at $1245$\,GHz, the variance from CIB anisotropies (without cleaning) yields $2 \sigma_{\bar S b, {\rm CII}} = 3.6\kjysr$, a factor of $34$ over instrumental noise on $f_{\rm sky} = 0.18$. Model the CIB in all bands in the data cube with the \citet{2012ApJ...752..120A} SED at each frequency, initially assuming full coherence between frequencies and neglecting correlation with the line intensity. After applying two-channel cleaning as in Sec.\,\ref{ss:galcont}, residual CIB anisotropies increase $\sigma_{\bar S b, {\rm CII}}$ by $10\%$ over instrumental noise. The CIB from veto channels is not fully spatially incoherent with CIB emission in the central band. For example, \citet{2017MNRAS.466..286M} find a correlation of $0.949$ between the CIB power spectra at $545$\,GHz and $857$\,GHz. Adding this level of decorrelation to the CIB realization in the two veto bands (which is conservative due to the proximity of $1185$, $1245$ and $1305$\,GHz), errors on $\sbcii$ after cleaning are double those from instrumental noise.

Dusty extragalactic point sources also contribute continuum shot noise. Assume that point sources are not masked or otherwise subtracted from the maps. Use the $857$\,GHz $dN/dS$ law from \citet{2017MNRAS.466..286M} and integrate $S^2 dN/dS$ to the maximum source flux $151$\,Jy \citep{2016A&A...594A..26P}. This gives $C_\ell = 20400\,{\rm Jy}^2/{\rm sr}$, which is a factor of $2.8$ larger than the shot-noise power spectrum in \citet{2017MNRAS.466..286M}, who mask bright sources. When smoothed at the PIXIE beam scale, dusty point sources contribute surface brightness fluctuations that have $7\%$ of the rms of the clustered CIB. Additional cleaning may be possible by extrapolating fluxes from a higher-resolution survey and subtracting the response to each source, or through masking of bright sources.

\subsection{Correlated continuum emission}
\label{ss:corrcib}

\begin{figure}
\epsscale{1.2}
\plotone{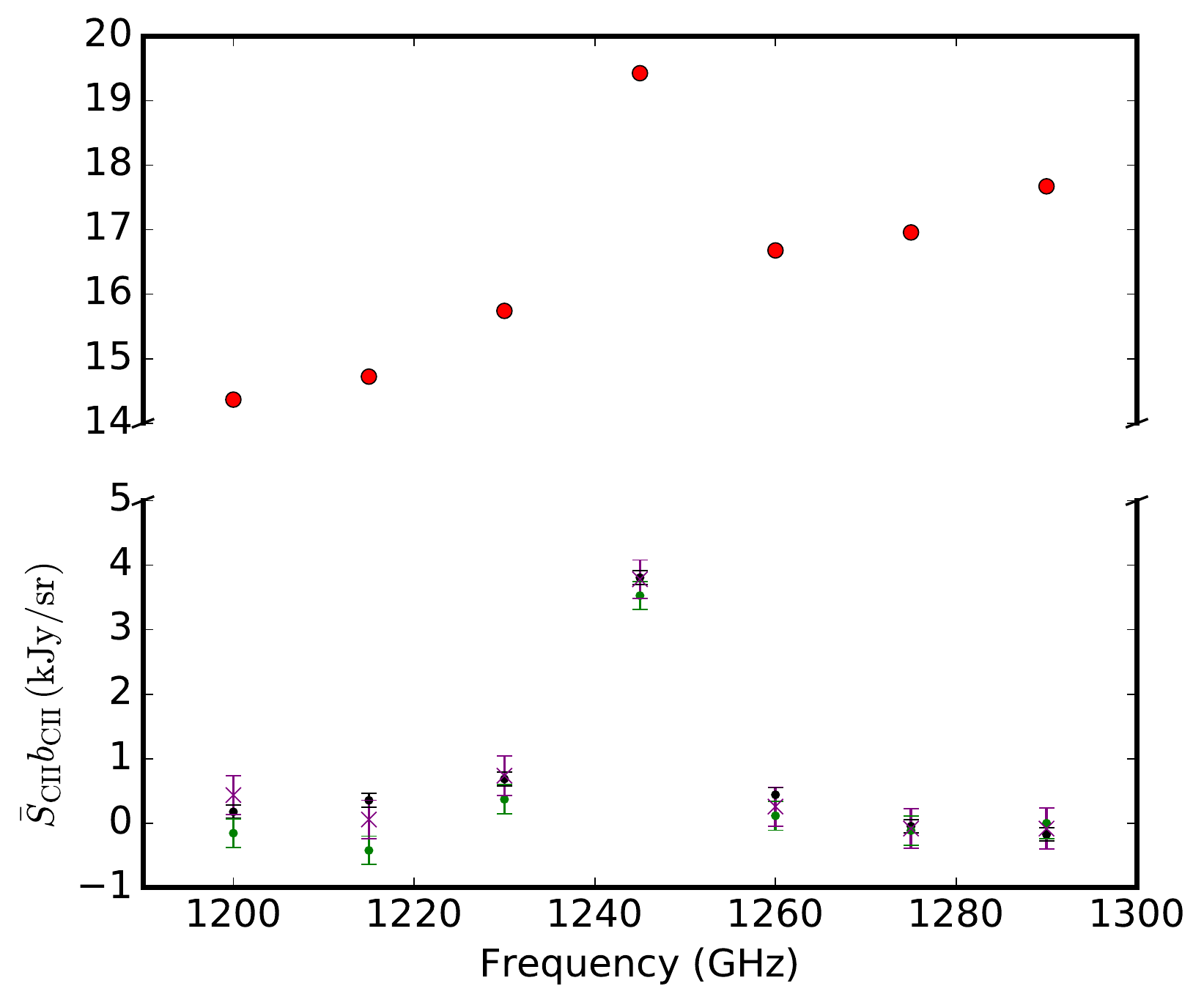}
\caption{Simulated constraints on $\sbcii$ at $1245$\,GHz from the cross-power, using Eq.\,\ref{eq:lbyl} and Eq.\,\ref{eqn:perlvar} for errors. Cosmological line signal in the intensity cube is the observed BOSS CMASS-North overdensity multiplied by $\sbcii = 4\kjysr$ (taking $\bar S_{CII}=2\kjysr$ and $b_{CII}=2$), convolved by the PIXIE beam, and with added instrumental noise. At each frequency, $\sbcii$ is estimated using the cross-correlation with BOSS data binned in the $1245$\,GHz slice. For example, the points at $1260$\,GHz are based on the cross-correlation of the intensity map at $1260$\,GHz and galaxy overdensity binned into redshifts consistent with \cii\ in the $1245$\,GHz channel. Black points show $\sbcii$ for signal and instrumental noise. Only PIXIE data at $1245$\,GHz correlate strongly with BOSS binned into $1245$\,GHz, and the input value of $4\kjysr$ is recovered well. Red points add coherent CIB produced by the overdensity at $1245$\,GHz. Correlated CIB emission biases all values of $\sbcii$ here due to the spectrally smooth thermal emission of dust. Red points have no errors indicated because that cross-power is necessarily performed before continuum cleaning and errors are not meaningful. Green points add galactic foregrounds and (dominant) uncorrelated CIB and then clean continuum emission using the linear combination of maps at $1185$ and $1305$\,GHz. Purple points remove the bias from residual correlated foregrounds.}
\label{fig:corrcib}
\end{figure}

The surface brightness of thermal dust emission is also modulated by overdensity \citep{2014A&A...570A..98S} and will produce a cross-correlation with the galaxy redshift survey.  For purposes here, calculations make several approximations (with incurred errors indicated) to provide an intuitive model for the interaction of correlated foregrounds with cleaning and the cross-power measurement. The correlation of the dust continuum with overdensity, across all pairs of frequency and galaxy survey redshift, is interesting in its own right as a decomposition of the CIB and its SED as a function of redshift \citep{2014A&A...570A..98S}, and warrants future simulation work. The first approximation is to describe the CIB through its emissivity density using the model and parameters of \citet{2010ApJ...718..632H} rather than through the luminosity function and underlying halo model \citep{2012MNRAS.421.2832S}. The Limber integral then gives the three two-point correlations $\{ C^{CIB \times g}_\ell, C^{gal}_\ell, C^{CIB}_\ell\}$ between the CIB and the galaxy survey binned onto a slice. For $C^{CIB \times g}_\ell$ \citep{2014A&A...570A..98S},
\begin{equation}
C^{CIB \times g}_\ell = \int \frac{dz}{\chi^2} \left ( \frac{d \chi}{d z} \right )^{-1} b_g b_{\rm CIB}(k, z) \frac{dN}{dz} \frac{dS_\nu}{dz} P_{\delta \delta} (k, z),
\label{eqn:corrciblimber}
\end{equation}
where $\chi(z)$ is the comoving distance, $b_{\rm CIB}(k,z)$ is the bias of dust emission, $dN/dz$ is the galaxy selection function, $dS_\nu/dz$ is the distribution of CIB emission \citep{2010ApJ...718..632H}, and $P_{\delta \delta} (k, z)$ is the underlying dark matter power spectrum evaluated at $k=\ell / \chi$. The Limber approximation recovers the full projection integral \citep{2013JCAP...11..044D} for the matter auto-power in the $1245$\,GHz channel to $5\%$ accuracy in rms map fluctuations smoothed on the PIXIE beam scale. We initially take constant, scale-independent bias for the CIB. From the Limber integrals, calculate the stochasticity $r_\ell^{CIB \times g} = C_\ell^{CIB \times g} / \sqrt{C_\ell^{CIB} C_\ell^{gal}}$. Multiplicative factors such as constant bias and brightness factor out. The correlated part of the intensity map can then be drawn from the power spectrum $(r_\ell^{CIB \times g})^2 C_\ell^{\rm CIB, Mak}$, which scales the total CIB anisotropy empirical model of Eq.\,\ref{eqn:makcib} \citep{2017MNRAS.466..286M} by the fraction of variance correlated with the galaxy overdensity in the intensity slice. This prescription reproduces the cross-power in \citet{2014A&A...570A..98S} to within $20\%$.

The signal model in simulations here is the actual BOSS-CMASS data binned onto PIXIE's bands, so the CIB realization must correlate specifically with BOSS. To that end, approximate the correlated CIB as a simple amplitude of CIB surface brightness produced at mean density times the overdensity in that slice $\bar S_{\rm CIB}^{\rm corr}(\nu_i) {\boldsymbol \delta} (\nu_i)$. At $1245$\,GHz and at PIXIE's resolution, $\bar S_{\rm CIB}^{\rm corr} (\nu_i) = 16\kjysr$. As an alternative to a constant bias $b_{\rm CIB}$ (which factors out of $r_\ell^{CIB \times g}$), the scale- and redshift-dependent bias model and estimated parameters in \citet{2013MNRAS.436.1896A} gives $\bar S_{\rm CIB}^{\rm corr}(\nu_i)$ different by $3\%$. The simple scaling from overdensity to the correlated CIB by $\bar S_{\rm CIB}^{\rm corr}(\nu_i)$ neglects differences in scale dependence ($<30\%$ in the relevant low-$\ell$ range) of the CIB and overdensity. In this formulation, both $\sbcii$ and $\bar S_{\rm CIB}^{\rm corr}$ directly multiply ${\boldsymbol \delta}$, so the cross-power measurement of $\sbcii$ is simple and additively biased by the thermal emission at mean density, as $\sbcii + \bar S_{\rm CIB}^{\rm corr}$. The emission spectrum $dS_\nu/dz$ of \citet{2010ApJ...718..632H} in Eq.\,\ref{eqn:corrciblimber} determines the SED of the correlated CIB, which is emitted by a lower-redshift slab than the bulk CIB so rises and peaks at a higher frequency.

Fig.\,\ref{fig:corrcib} shows the inferred value of $\sbcii$ from mock cross-power measurements, including a correlated CIB contribution. The measurement of $\sbcii$ is based on the cross-power of the intensity survey at each frequency with galaxy overdensity data at fixed redshift corresponding to \cii\ at $\nu=1245$\,GHz. The correlated CIB is spectrally highly coherent, so it contributes to the cross-power in all frequency slices. In contrast, the line intensity signal correlation is negligible at channel separations greater than two bins, and dominates only at $\nu=1245$\,GHz. Hence continuum contamination that correlates with cosmological overdensity can be isolated as a plateau of correlation at offsets in frequency or redshift. Further, cleaning the total dust continuum emission suppresses correlated CIB contributions. Fig.\,\ref{fig:corrcib} shows results of a simple two-channel cleaning procedure. Non-zero correlation at offset channels can then provide evidence for residual correlated emission terms.

Extending the foreground cleaning expression in Sec.\,\ref{ss:galcont} to separate correlated and uncorrelated foregrounds, the cleaned map becomes $(1-{\mathbf \Pi}) ({\bf s}_\nu^{\rm signal} + {\bf s}_\nu^{\rm uncorr\,fg} + {\bf s}_\nu^{\rm corr\,fg})$. The linear combination parameters in ${\mathbf \Pi}$ primarily minimize variance from galactic dust. While much of the CIB is also subtracted, it is imperfect because of the difference in SED. Some residuals of correlated CIB remain and are anticorrelated with the overdensity (Fig.\,\ref{fig:corrcib}). A simple approach to correcting this residual bias exploits the spectrally smooth nature of the CIB by fitting a baseline at frequency offsets (here, $\{ 1200, 1215, 1275, 1290 \}$\,GHz) for which line signal correlations are small and non-zero correlation signal is produced by residual continuum emission. Simulations over the narrow spectral range here use a linear baseline fit and propagate errors to the final estimate.

An alternative approach estimates $\bar S_{\rm CIB}^{\rm corr}$ and $\sbcii$ jointly using the likelihood of the cross-powers at all frequency offsets and at each redshift of the galaxy survey. This constrains the correlated CIB model in parallel with $\sbcii$. This is roughly related to fitting a continuum baseline in the red points of Fig.\,\ref{fig:corrcib}. However, in this measurement, no dust continuum cleaning could be performed because that operation would also clean the CIB. This produces higher variance estimates and a complex likelihood, but may warrant future studies. If $\bar S_{\rm CIB}^{\rm corr}$ estimates exist from a model, they can be combined with ${\boldsymbol \delta}$ from the galaxy survey to estimate biases. 

\subsection{Contamination at $\nu < 600$\, GHz}

Galactic foregrounds from dust, synchrotron, spinning dust and free-free emission all become relevant in the regime of CMB spectral distortions (taken here as $\nu < 600$\, GHz). CMB and thermal Sunyaev-Zel'dovich \citep[tSZ,][]{1969Ap&SS...4..301Z} anisotropies add variance in addition to CIB described previously. Take a band at $165$\,GHz that is near the maximum amplitude of the CMB and maximum decrement of tSZ. Again {\tt PySM} \citep{2016arXiv160802841T} provides a model of galactic foregrounds listed above. For tSZ, simulations use a realization of the thermal Compton parameter from the power spectrum estimated in \citet{2016A&A...594A..22P}, scaled to surface brightness at each frequency. At $165$\,GHz and outside of the galactic plane, CMB anisotropy dominates and has rms $25\kjysr$ fluctuations (convolved on the beam scale), while the CMB monopole is $383\mjysr$. For comparison, the CMB monopole surface brightness at $1245$\,GHz is $0.75\kjysr$, and so was neglected earlier. Hence, compared to simulations at $1245$\,GHz described earlier, the contaminating anisotropy is much lower and the overall surface brightness is much higher.

To the extent that the calibrator blackbody spectrum matches the CMB, the differencing FTS cancels the CMB monopole. The calibration requirements from Sec.\,\ref{ss:instfg} then split into the spectral stability of (1) the nulling reference and (2) the instrumental response characterization, which applies after nulling. Again spectral stability is the central quantity and must be controlled to $<1 \times 10^{-6}$ (which is compatible with the $3 \times 10^{-7}$ physical blackness of the calibrator).

We use two-channel cleaning based on maps at $135$\,GHz and $195$\,GHz to clean the central $165$\,GHz band. Rather than \cii, this section uses CO $2-1$ as a representative transition. From Eq.\,\ref{eqn:perlvar}, the overall sensitivity depends on the noise weighted by $C_\ell^{\delta \delta}$. Hence the relative impact of residual foregrounds has some scale dependence. A low-redshift line has more structure at low-$\ell$ and $\sigma_{\bar S b}$ will be more sensitive to the larger residual foregrounds there (Fig.\,\ref{fig:galactic_resid}). Calculations here continue to use $z=0.4$ (a relatively low redshift) to be conservative. CMB anisotropy alone is removed up to a $28\%$ increase in $\sigma_{\bar S b}$, which is dominated by instrumental noise in the linear combination of bands ($3\%$ from residual CMB variance). At $165$\,GHz, contaminants have a range of spectral characters. The linear combination of $135$ and $195$\,GHz has two degrees of freedom, and primarily cleans the CMB (a factor of $270$ suppression in rms). However, dusty components are still suppressed by a factor of approximately $\approx 6$ and the SZ is suppressed by a factor of $10$ (in rms). On the cleanest $25\%$ of the sky, galactic foregrounds yield residuals in the two-band linear combination cleaning that boost $\sigma_{\bar S b}$ by $27\%$ over the contribution from instrumental noise. Residual contamination from tSZ and CIB contributes $2\%$ and $7\%$ additionally. An alternative to the tSZ power spectrum is to directly use the tSZ map estimated in \citet{2016A&A...594A..22P}. This is $1.6$ times larger in rms on the PIXIE beam scale, but also contributes negligible residuals. As in Sec.\,\ref{ss:corrcib}, tSZ correlates with the overdensity and is spectrally smooth, so a similar offset frequency test or baseline subtraction apply (as needed). Further reduction in residuals could be achieved by reconstructing high signal-to-noise CMB anisotropy and tSZ templates using all PIXIE bands.

\section{Summary}
\label{sec:summary}

Viewed as a tomographic measurement, deep surveys for CMB spectral distortions also contain redshifted CO and \cii\ line emission. The ratio of the SFR to cold gas is related to the star-formation efficiency, making a survey of both \cii\ and CO compelling. CMB spectrometer architectures such as PIXIE probe large, linear scales. These scales have lower variance and fewer available modes, but they directly trace $\sbgas$ without the complications of nonlinear evolution, halo occupation and stochasticity. Detection through cross-correlation evades cosmic variance and provides an unambiguous detection over residual continuum contamination and interlopers. 

A simple two-channel linear combination cleaning approach removes much of the contamination in simulations at $1245$\,GHz and $165$\,GHz, approximately representative of maximum thermal dust and CMB emission, respectively. $\sbcii$ is recovered robustly in intensity data cubes with galactic and correlated plus uncorrelated extragalactic emission. The cleaning demonstrated here suggests that a joint map-space likelihood of templates and galaxy overdensity maps could extract line brightness as a function of redshift. Uncertainty in the bias of the galaxy sample propagates directly into $\sbgas$ estimates, and must be included as a prior.

Future work must consider the interpretation of $\sbgas$ from \cii\ and CO (including from higher $J$) in terms of galaxy evolution, accounting for bias. Related studies of $\Omega_{\rm HI}$ \citep[e.g.][]{2016MNRAS.458..781P, 2016MNRAS.456.3553V} have jointly analyzed $b_{\rm HI} \Omega_{\rm HI}(z)$ from intensity mapping with $\Omega_{\rm HI}(z)$ from studies of individual objects. Sensitivity to the line luminosity-weighted bias is a unique strength of intensity mapping. This bias, in turn, connects the processes of star formation to the cosmological setting and host halos. In addition to the angular power spectra described here, $P(k_\perp, k_\parallel)$ has the promise of separating brightness and bias through redshift-space distortions \citep[e.g.][]{2010PhRvD..81f2001M}.

\section{Acknowledgements}
\label{sec:ack}

I thank Alan Kogut and Dale Fixsen for comments and discussion of PIXIE and FIRAS, and for providing monopole and per-pixel sensitivity for PIXIE, and Natalie Mashian for providing a CO cumulative emission model. I acknowledge the organizers and participants of the Stanford/SLAC intensity mapping workshop for stimulating discussion, and comments from Adam Lidz, David Spergel, and an anonymous reviewer.

\bibliographystyle{apj}
\bibliography{cold_gas_pixie}

\end{document}